# HIGGS BOSON PRODUCTION
# AND WEAK BOSON STRUCTURE[*]

Wojciech SŁOMIŃSKI and Jerzy SZWED

Institute of Computer Science, Jagellonian University,
Reymonta 4, 30-059 Kraków, Poland

## Abstract

The influence of the QCD structure of the weak bosons on the Higgs boson production in $e$-$p$ scattering is studied. The energy and Higgs boson mass dependence of the cross-section, following from the new contributions, is calculated.

---

[*]Work supported by the Polish State Committee for Scientific Research (grant No. 2 PO3B 081 09) and the Volkswagen-Stiftung.

# 1 Introduction

In our recent papers [1, 2] we have introduced the basic properties of the W and Z boson structure functions. In analogy to the photon case [3] it has been shown there, how the quark and gluon content of the intermediate bosons appears due to the QCD cascade. The corresponding evolution equations have been solved in the asymptotic regime with the solutions developing logarithmic $Q^2$ growth of the quark and gluon densities. The form of weak couplings forces these densities to depend strongly on spin and flavour.

It is extremely interesting where the QCD structure of weak intermediate bosons may be observed experimentally. In the systematic investigation of possible processes we start with the Higgs boson production. The question put forward in this case is to what extent the inclusion of the 'resolved' W and Z can modify the predictions known before. We concentrate on the Higgs boson production in $e$-$p$ scattering, quoting the $e^+e^-$ scattering, where one should not expect large contributions, for completeness.

The paper is organized as follows. In Section 2 we recall the formalism introduced before to study the QCD structure of gauge bosons and quote the main results concerning W and Z, compared to the photon case. Section 3 presents the calculation of the cross-section for the Higgs boson production in $e$-$p$ scattering at selected energies. The results are compared with the dominant production channel [4, 5] (W-W fusion) and earlier calculation of the 'resolved' photon contribution [6]. Summary, comments and conclusions are given in Section 4.

# 2 QCD structure of $W$ and $Z$ bosons

In the standard model weak intermediate bosons are elementary (point-like) particles. Nevertheless, when observed by a very high $Q^2$ particle, they can reveal QCD structure by collinear quark-gluon Bremsstrahlung. In this sense they become "composite".

The evolution equation for unpolarized parton $\mathcal{A}$ density, $f_\mathcal{A}^B(x,t)$, inside a composite weak intermediate boson $B$ reads

$$\frac{df_\mathcal{A}^B(x,t)}{dt} = \sum_\mathcal{B} \mathcal{P}_{\mathcal{AB}}(x,t) \otimes f_\mathcal{B}^B(x,t), \qquad (1)$$

where $t = \ln(Q^2/Q_0^2)$ and the scale $Q_0^2$, which depends on the particular process, will be discussed later. $\mathcal{P}_{\mathcal{AB}}(x;t)$ are splitting functions and the convolution is defined as

$$(P \otimes f)(x) \equiv \int dx_1\, dx_2\, P(x_1)\, f(x_2)\, \delta(x - x_1 x_2). \qquad (2)$$

For weak bosons the indices $\mathcal{A}, \mathcal{B}$ go over quarks, antiquarks, gluons and point-like $\gamma$, W and Z.

In the following we will consider the leading-log QCD and 1-st order electroweak case. Denoting weak intermediate bosons by upper case letters and QCD partons by lower case ones, we have

$$f_{AB}(x,t) = \delta_{AB}\delta(1-x), \qquad (3)$$

$$\mathcal{P}_{iB}(x,t) \equiv \frac{\alpha_{\text{em}}}{2\pi} P_{iB}(x), \qquad (4)$$

$$\mathcal{P}_{ik}(x,t) \equiv \frac{\alpha_{\text{s}}(t)}{2\pi} P_{ik}(x). \qquad (5)$$

Substituting this into Eq.(1) we arrive at the following non-homogeneous evolution equations for the QCD content of a weak intermediate boson:

$$\frac{df_i^B(x,t)}{dt} = \frac{\alpha_{\rm em}}{2\pi} P_i^B(x) + \frac{\alpha_{\rm s}(t)}{2\pi} \sum_k P_{ik}(x) \otimes f_k^B(x,t) \tag{6}$$

.

In the lowest $\alpha_s$ order the splitting functions of the longitudinal $W$ and $Z$ bosons vanish and the transverse ones read

$$P_{{\rm q}_\pm Z_\perp}(x) = P_{\bar{\rm q}_\pm Z_\perp}(x) = z_{{\rm q}_\pm} \, s(x), \tag{7}$$

$$P_{{\rm d}_- W_\perp^-}(x) = P_{\bar{\rm u}_+ W_\perp^-}(x) = \frac{1}{2\sin^2\theta_{\rm W}} \, s(x), \tag{8}$$

with $s(x) = [x^2 + (1-x)^2]/2$.

The QCD splitting functions are taken in the standard form [7].

In the leading-log approximation with $\alpha_{\rm s}(t)$

$$\alpha_{\rm s}(t) = \frac{2\pi}{bt}, \tag{9}$$

($b = 11/2 - n_{\rm f}/3$ for $n_{\rm f}$ flavours) the t dependence of the equations (6) can be factorized out

$$f_k^B(x,t) \simeq \frac{\alpha_{\rm em}}{2\pi} \tilde{f}_k^B(x) t \tag{10}$$

leaving integral equations for the $x$ dependence

$$\tilde{f}_i^B(x) = P_i^B(x) + \frac{1}{b} \sum_{k={\rm q},\bar{\rm q},G} P_{ik}(x) \otimes \tilde{f}_k^B(x), \tag{11}$$

Numerical solutions to the above equations can be found in Ref. [2]. They show in general that, apart from the $t$-dependent factor which at sub-asymptotic momentum transfers might be different (see discussion in the next Section), the quark and gluon structure of the weak bosons is reacher than that of the photon.

In the above review we have summed over the parton and weak boson polarizations (leaving out the longitudinal bosons which do not contribute in leading-log approximation). The same considerations can be repeated keeping the parton and boson polarizations fixed [8]. In fact, due to the particular form of weak couplings, we shall use in the following calculations partonic densities of given polarization inside polarized $W$ and $Z$ bosons.

## 3  Higgs production in $e$-$p$ scattering

The dominant mechanism of the Higgs boson production in electron–proton scattering is the $W$-$W$ and $Z$-$Z$ fusion (see Fig. 1a) [4, 5]. The new diagrams, which involve gauge boson structure, are shown in Fig. 1b–c. The case with the 'resolved' photon has been already studied some time ago [6]. Here we include in the calculation the $W$ and $Z$ bosons.

The approximations made in the following considerations require a word of warning. First, we will be using the equivalent boson approximation, known since long for the photon [9], and introduced in Ref. [10] for the $W$ and $Z$. In the case of massive weak bosons, out of the three possible polarization states only the transverse degrees of freedom develop the logarithmic factor, characteristic of photon emission. The density of transversely polarized gauge bosons inside unpolarized electron

$$f_B^e(x) \propto \frac{1}{x} \ln\left(\frac{Q_{\max}^2 + M_B^2}{Q_{\min}^2 + M_B^2}\right) \qquad (12)$$

with $x$ — the boson momentum fraction, $M_B$ — the gauge boson mass and $Q^2$ — the negative momentum squared of the emitted gauge boson. One sees that whereas in the photon case the logarithm is scaled by the electron mass, coming from $Q_{\min}^2$, in the weak sector it is the mass $M_B^2$ which sets up the scale (explicit forms used in the considered processes are given in the next Section). Consequently, due to the large weak boson mass, the above logarithmic factor is responsible at presently available energies for the fact that there is more 'equivalent' photons in the electron than $W$'s and $Z$'s. The accuracy of the equivalent boson approximation has been tested in the processes where exact calculation is also possible [5]. For example the ratio of the approximate to exact results in the Higgs production in $e^+e^-$ scattering varies between a factor of two and few percent in the energy range between 500 GeV and 50 TeV and approaches 1 with increasing Higgs boson mass. Taking into account the above remarks one should treat the numerical results at energies corresponding to the existing accelerators only as estimates.

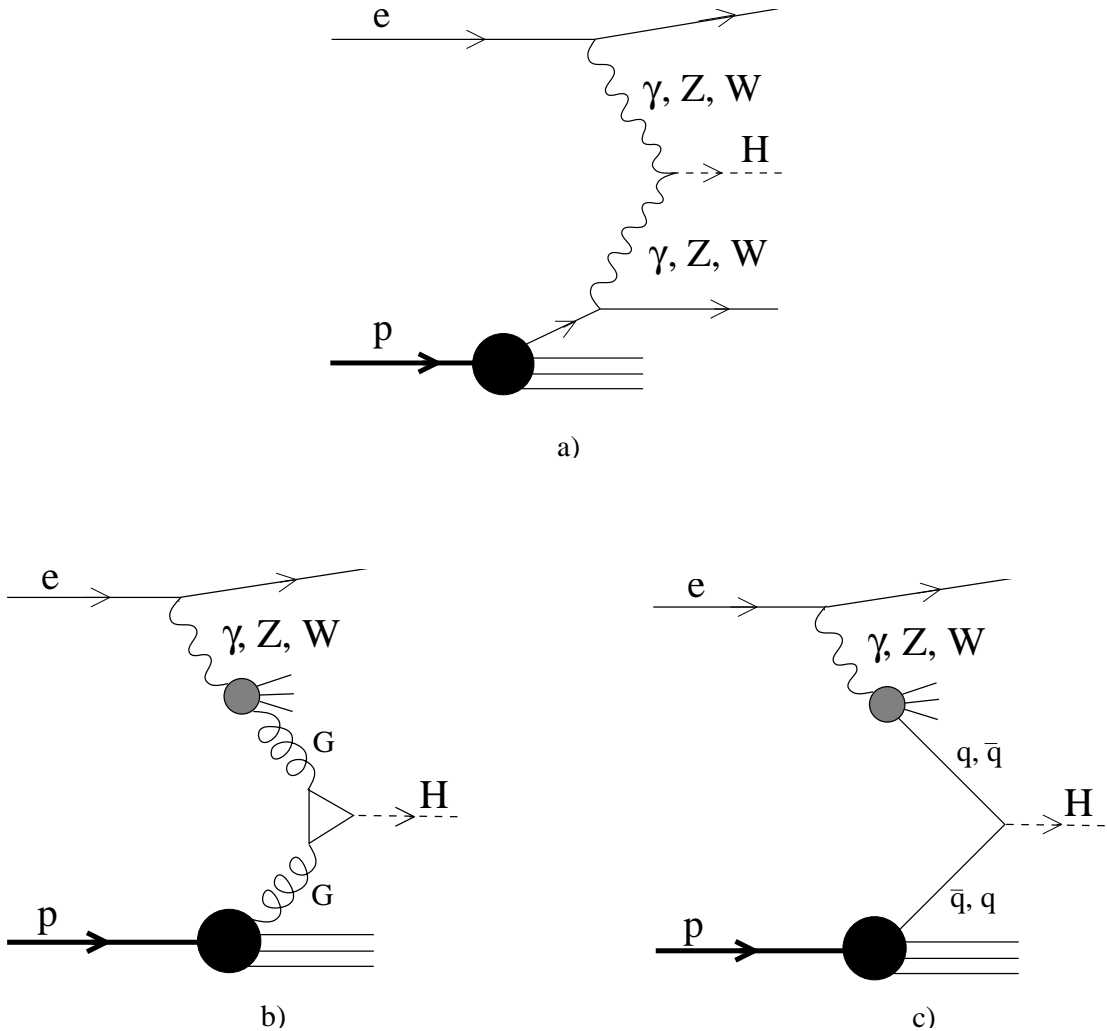

Figure 1: Diagrams contributing to the Higgs boson production in e-p scattering: a) dominant W-W fusion; b) gluonic part of the electroweak boson structure; c) quark part of the electroweak boson structure.

Another problem is the $Z$-$\gamma$ interference. In general the neutral current exchange contains a coherent mixture of the $Z$ and the photon. However in the probabilistic approach used here these possible interference terms are neglected.

The next approximation concerns the quark masses. In the QCD evolution equation all masses are neglected, they are used solely as thresholds, opening new flavour evolution when increasing $Q^2$. A more delicate treatment requires the $b$ quark threshold in the $W$ structure evolution. In principle it appears already above the $b$ quark mass due to the $b\bar{u}$ production. However only after the channel $b\bar{t}$ opens ($\sqrt{Q^2} > 180$ GeV), the Kobayashi–Maskawa matrix elements squared add up to 1 and do not suppress the b quark production. In our calculation we neglect this suppression which potentially exists in the intermediate Higgs boson mass range.

Finally the scale $Q_0^2$ of the QCD evolution requires some attention. In the leading-log approximation, which we are using, its value for $Q^2 \gg Q_0^2$ is formally irrelevant (changing $Q_0$ gives next-to-leading-log corrections). At finite $Q^2$ the choice depends on the physical situation in the process. In $e$-$p$ scattering we are interested in the electroweak bosons, 'as few off shell as possible' which practically means the mass negative and close to zero. Therefore, as concerns the masses, the difference between photons and weak bosons vanishes and in all cases we should take $Q_0^2 = \Lambda_{\rm QCD}^2$.

The contribution to the total cross-section for the process $e^-p \to \nu/e^- H X$ coming from the 'resolved' bosons (Fig. 1 b,c) reads

$$\sigma^B = \sum_{i,j,\alpha,\beta} \int_\tau^1 \frac{dx}{x} f^{\rm e}_{B_\beta}(x) \int_{\tau/x}^1 \frac{dy}{y} f^{B_\beta}_{i_\alpha}(y) f^P_{j_\alpha}\left(\frac{\tau}{xy}\right) \hat{\sigma}_{ij} \qquad (13)$$

where $\tau = M_H^2/s$ with $M_H$ being the Higgs boson mass and $s$ — the total c.m. energy squared. The function $f^{\rm e}_{B_\beta}$ is the boson $B$ (of polarization $\beta$) density inside electron, $f^{B_\beta}_{i_\alpha}$ is the quark/antiquark/gluon (of polarization $\alpha$) density inside the boson $B$ (of polarization $\beta$) and $f^P_{j_\alpha}$ — the antiquark/quark/gluon (of polarization $\alpha$) density inside the proton. The sum extends over partons ($i,j = q, \bar{q}, G$), their polarization $\alpha$ and the boson polarization $\beta$. The polarization–independent partonic cross-section

$$\hat{\sigma}_{GG} = \frac{\pi}{144\sqrt{2}} \left(\frac{\alpha_{\rm s}}{\pi}\right)^2 |N|^2 \frac{G_{\rm F} M_H^2}{s} \qquad (14)$$

for gluon–gluon (Fig. 1b) and

$$\hat{\sigma}_{q\bar{q}} = \frac{\sqrt{2}\pi}{3} \frac{G_{\rm F} m_q^2}{s} \qquad (15)$$

for quark–antiquark annihilation (Fig.1c). Here $G_{\rm F}$ is the Fermi coupling, $m_q$ — the quark mass and $N$ — a function of the quark and Higgs boson masses [11].

The equivalent boson densities depend on the exchanged boson and its polarization:

$$f^{\rm e}_{W_\pm}(x) = \frac{\alpha}{4\pi} \left(w_\pm + w_\mp (1-x)^2\right) \frac{1}{x} \ln\left(\frac{xs + M_W^2}{M_W^2}\right), \qquad (16)$$

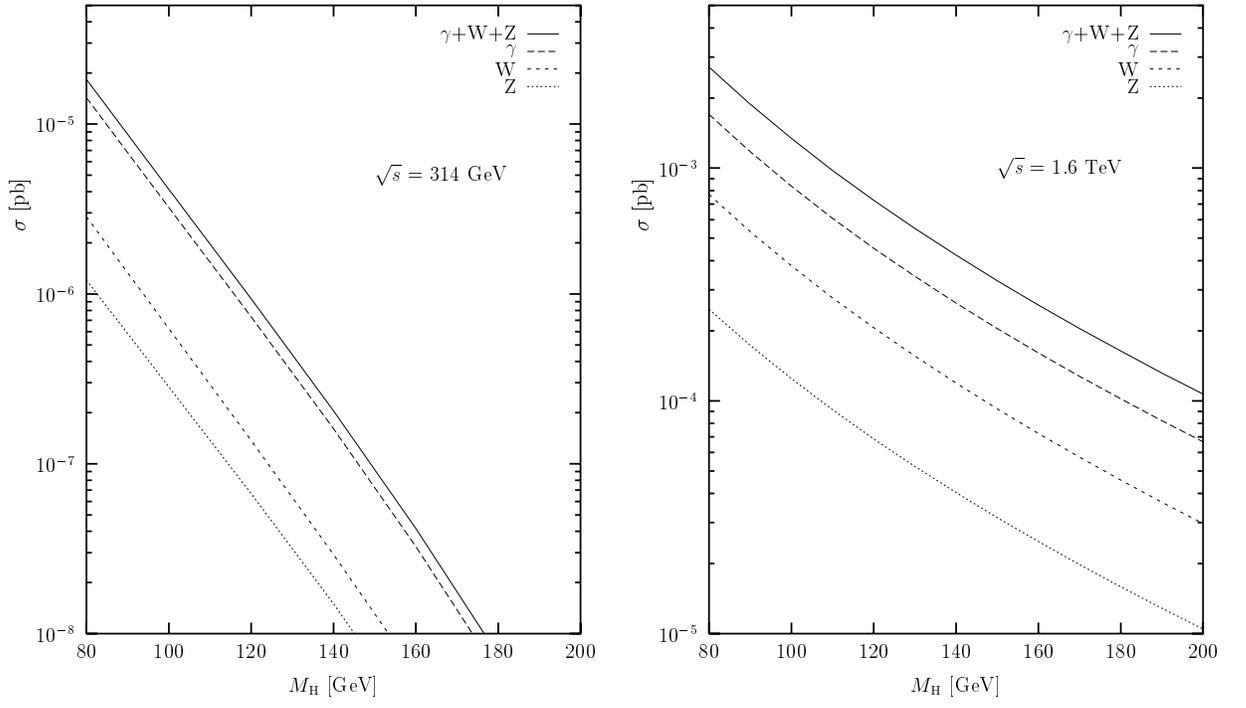

Figure 2: Contribution from the 'resolved' electroweak bosons to the cross-section for the Higgs boson production in $e$-$p$ scattering at: a) $\sqrt{s} = 314$ GeV and b) $\sqrt{s} = 1.6$ TeV as function of the Higgs boson mass.

$$f^{\mathrm{e}}_{Z_\pm}(x) = \frac{\alpha}{4\pi}\left(z_\pm + z_\mp(1-x)^2\right)\frac{1}{x}\ln\left(\frac{xs + M_Z^2}{M_Z^2}\right), \qquad (17)$$

$$f^{\mathrm{e}}_{\gamma_\pm}(x) = \frac{\alpha}{4\pi}\left(1 + (1-x)^2\right)\frac{1}{x}\ln\left(\frac{(1-x)s}{xm_{\mathrm{e}}}\right), \qquad (18)$$

where

$$w_+ = \frac{1}{2\sin^2\theta_{\mathrm{W}}},\ w_- = 0,\ z_+ = \tan^2\theta_{\mathrm{W}}\left(1 - \frac{1}{2\sin^2\theta_{\mathrm{W}}}\right)^2,\ z_- = \tan^2\theta_{\mathrm{W}} \qquad (19)$$

and $m_{\mathrm{e}}$ is the electron mass.

The parton densities $f^{B_\beta}_{i_\alpha}$ fulfill several relations [8], for example in the case of $W^-$:

$$f^{W^-}_{d_+} = f^{W^+}_{\bar{u}_-}, \qquad (20)$$

$$f^{W^-}_{d_-} = f^{W^+}_{\bar{u}_+}, \qquad (21)$$

$$f^{W^+}_{d_+} = f^{W^+}_{\bar{d}_+} = f^{W^-}_{\bar{d}_-} = f^{W^-}_{\bar{u}_-} = f^{W^-}_{u_-} = f^{W^+}_{u_+}, \qquad (22)$$

$$f^{W^+}_{d_-} = f^{W^+}_{\bar{d}_+} = f^{W^-}_{\bar{d}_+} = f^{W^-}_{\bar{u}_+} = f^{W^-}_{u_+} = f^{W^+}_{u_-}, \qquad (23)$$

$$f^{W^+}_{G_-} = f^{W^-}_{G_+}, \qquad (24)$$

$$f^{W^+}_{G_+} = f^{W^-}_{G_-}. \qquad (25)$$

In the examples shown below we use their explicit asymptotic form following from the numerical solutions of the Eqs.(11). Only in the case of the photon we are able to see whether the asymptotic solutions lead to different results than the more realistic parametrisations of its structure. We have checked that the parametrisation of Ref.[12] (LAC3) gives the cross-section for the $e$-$p$ Higgs boson production up to 30% larger in the considered energy and Higgs boson mass range.

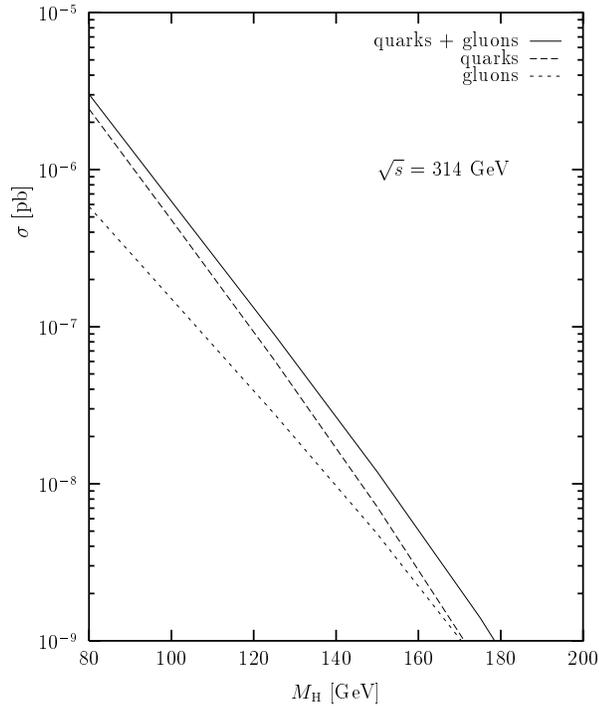

Figure 3: The 'resolved' $W$ contribution to the cross-section for the Higgs boson production in $e$-$p$ scattering at $\sqrt{s} = 314$ GeV as function of the Higgs mass. The gluonic and quark parts plotted separately.

The parton densities inside the proton $f_i^P(x)$ are taken from the parametrisation of Ref. [13] (MRS3).

The results for scattering energies $\sqrt{s} = 314$ and 1600 GeV are shown in Figs. 2 and 3. One sees that the 'resolved' $W$ contribution is about a factor of 4 smaller than that of the 'resolved' $\gamma$ at lower energies $\sqrt{s} = 314$ GeV and approaches one half at $\sqrt{s} = 1.6$ TeV (Fig.2). The $Z$ contribution is the smallest for all considered energies and Higgs boson masses. In all cases the quark–antiquark diagram (Fig. 1c) is larger than the gluonic one (Fig. 1b), mainly due to presence of the $b$ quark (Fig 3.). One should keep in mind that the dominant term in the Higgs production cross-section, the $W$-$W$ fusion (Fig. 1a) [4, 5] exceeds the 'resolved' boson contribution by at least an order of magnitude. The conclusion is rather obvious: the QCD structure of the gauge bosons, used in $e$-$p$ scattering, does not help in the hunt for the Higgs boson.

## 4 Summary

In the paper we have considered the influence of the QCD structure of the electroweak gauge bosons on the $e$-$p$ production of the Higgs boson. We have found that using the asymptotic form of the 'resolved' W and Z, the contribution to the cross-section of the W structure is of the same order as that of the photon, the Z term being slightly smaller. In general however the considered new diagrams cannot compete with the dominant channel i.e. the W-W fusion.

We have also checked the Higgs boson production via 'resolved' bosons in $e^+e^+$ scattering. As expected, the cross-section is suppressed additionally, as compared to the W-W and Z-Z fusion, by a factor of $\alpha^2$ which multiplies the parton densities and consequently is negligible.

One general lesson follows from the above studies. The W structure function contributes to the considered processes approximately with the same strength as the structure of the photon. This means that one should look for its appearance in the reactions where the 'resolved' photon is known to dominate.

# 5  Acknowledgments

This work has been performed during our visit to DESY, Hamburg. We would like to thank the DESY Theory Group for hospitality and the Volkswagen Foundation for financial support.

# References

[1] W. Słomiński and J. Szwed, Phys. Lett. **B** 323 (1994) 427.
    W. Słomiński and J. Szwed, in *High Energy Spin Physics*, Proc. of X Int. Symp. on high Energy Spin Physics, Nagoya, Japan, 1992; ed. T. Hasegawa et al. (Universal Academy, Tokyo, 1993);

[2] W. Słomiński and J. Szwed, Phys. Rev. **D52** (1995).

[3] E. Witten, Nucl. Phys. **B120**, 189 (1977); C.H. Llewellyn Smith, Phys. Lett. **79B**, 83 (1978); R.J. DeWitt et al., Phys. Rev. **D19**, 2046 (1979); T.F. Walsh and P. Zerwas, Phys. Lett. **36 B** (1973) 195; R.L. Kingsley, Nucl. Phys. **B 60** (1973) 45.

[4] Z. Hioki et al., Progr. Theor. Phys **69** (1983) 1484; D.A. Dicus and S.D. Willenbrock, Phys. Rev. **D 32** (1985) 1642.

[5] G. Altarelli, B. Mele and F. Pitolli, Nucl. Phys. **B 287** (1987) 205.

[6] W. Słomiński and J. Szwed, Acta Physica Polonica **B 22** (1991) 859.

[7] G. Altarelli and G. Parisi, Nucl. Phys. **B126** (1977) 298.

[8] W. Słomiński and J. Szwed, unpublished.

[9] C. Weizsäcker and E.J. Williams, Z. Phys. **88**, 612 1934).

[10] G.L. Kane, W.W. Repko and W.B. Rolnick, Phys. Lett., **148B**, 367 (1984).

[11] R.N. Cahn and S. Dawson, Phys. Lett. **136 B** (1984) 196.

[12] H. Abramowicz, K. Charchula and A. Levy, Phys. Lett. **B 269** (1991) 458.

[13] A.D. Martin, W.J. Stirling and R.G. Roberts, Phys. Rev. **D 47** 867.